\documentclass[twocolumn,aps,showpacs,showkeys,preprintnumbers]{revtex4}                                   %two-column publication format
\usepackage{graphicx}
\usepackage{dcolumn}
\usepackage{amsmath}
\usepackage{amssymb}
\begin{document}
%\input{psfig}
%
%+++++++++++++++++++++++++++++++++++++++++++++++++++++++++++++++++++++++++++++
%
%  Macro definitions
%
%+++++++++++++++++++++++++++++++++++++++++++++++++++++++++++++++++++++++++++++
\def\etal{{\it et al~}}
%
%+++++++++++++++++++++++++++++++++++++++++++++++++++++++++++++++++++++++++++++
%
% Title of the paper
%
%+++++++++++++++++++++++++++++++++++++++++++++++++++++++++++++++++++++++++++++
%
\title{{\it K}-shell x-ray spectroscopy of atomic nitrogen}
%
% repeat the \author\affiliation pair as needed

\author{M. M. Sant'Anna}
\altaffiliation[Present address: ]{Instituto de F\'\i sica, Universidade Federal do Rio de Janeiro,
                                    Caixa Postal 68528, 21941-972 Rio de Janeiro RJ, Brazil}

\author{A. S. Schlachter}
\affiliation{Advanced Light Source, Lawrence Berkeley National Laboratory,
                   University of California, Berkeley, California 94720, USA}

\author{ G. \"{O}hrwall}
\altaffiliation[Present address: ]{MAX-Lab, Lund University, Box 118, Lund SE-221 00, Sweden}
\author{W. C. Stolte}
\author{D. W. Lindle}
\affiliation{Department of Chemistry, University of Nevada,
             Las Vegas, Nevada 89154-4003, USA}
	 	
\author{B. M. McLaughlin}
\altaffiliation[Present address: ]{School of Mathematics \& Physics,
                     Queen's University of Belfast, Belfast BT7 1NN, UK}
\affiliation{ITAMP, Harvard Smithsonian Center for Astrophysics, Mail Stop 14,
             60 Garden Street, Cambridge, Massachusetts 2138, USA}

\date{\today}
%
%+++++++++++++++++++++++++++++++++++++++++++++++++++++++++++++++++++++++++++++
%
%              Abstract
%
%+++++++++++++++++++++++++++++++++++++++++++++++++++++++++++++++++++++++++++++
\begin{abstract}
Absolute {\it K}-shell photoionization cross sections for atomic nitrogen
have been obtained from both experiment and state-of-the-art theoretical techniques.
Because of the difficulty of creating a target of neutral atomic nitrogen,
no high-resolution {\it K}-edge spectroscopy measurements have been reported
for this important atom. Interplay between theory and experiment enabled
identification and characterization of the strong $1s$ $\rightarrow$ $np$ resonance
features throughout the threshold region. An experimental value of 409.64 $\pm$ 0.02 eV
was determined for the {\it K}-shell binding energy.
\end{abstract}
\maketitle
%
%
%++++++++++++++++++++++++++++++++++++++++++++++++++++++++++++++++++++++++++++
%
%      Text of paper follows
%
%++++++++++++++++++++++++++++++++++++++++++++++++++++++++++++++++++++++++++++
The atomic form of the seventh element, nitrogen, plays important roles in such diverse areas
as x-ray astronomy, planetary physics, and materials science, fields that collectively span
about 28 orders of magnitude on the length scale. Yet, to date, the {\it K}-shell spectroscopy
of neutral atomic nitrogen has been the subject of limited theoretical predictions,
untested by any high-quality experimental measurements. It is the purpose of this work to
remedy this situation by reporting on high-resolution {\it K}-edge spectra, along with
state-of-the-art theory, providing absolute atomic nitrogen {\it K}-shell cross sections for the first time.

In x-ray astronomy, the satellites {\it Chandra} and {\it XMM-Newton} are providing a wealth of
x-ray spectra of astronomical objects. Spectroscopy in the soft-x-ray region (5-45 \AA),
including {\it K}-shell transitions of C, N, O, Ne, S, and Si and {\it L}-shell transitions of Fe and Ni,
provides a valuable tool for probing extreme environments in active galactic nuclei,
x-ray binary systems, cataclysmic variable stars, and Wolf-Rayet Stars \cite{wolf03,Skinner10},
as well as interstellar media (ISM) \cite{garcia11}. The latter, recent work, for example,
demonstrated that x-ray spectra from {\it XMM-Newton} can be used to characterize ISM,
provided accurate atomic oxygen {\it K}-edge cross sections are available.
Analogous results concerning the chemical composition of ISM are expected
with the availability of accurate atomic nitrogen {\it K}-edge cross sections \cite{garcia}.

In planetary science, inner-shell photoabsorption of atomic nitrogen is known
to affect the energetics of the terrestrial upper atmosphere and, together with atomic oxygen
and molecular nitrogen determine the ion-neutral chemistry and the temperature structure
of the thermosphere through production of nitric oxide \cite{link99}. Since the nitric-oxide
abundance is highly correlated with soft-x-ray irradiance, a full picture requires
accurate knowledge of both the solar flux and the x-ray cross sections of these species.
As demonstrated by observations using {\it Chandra}, atomic nitrogen and oxygen
 play a vital role in the terrestrial aurora, as well as similar phenomena on other planets,
 such as Jupiter \cite{bhardwaj07}.

In materials science, one application of {\it K}-edge spectroscopy of atomic nitrogen
is in determining how the surface topology of a crystal affects dissociation of nitric oxide molecules
adsorbed on stepped surfaces by measuring chemical shifts in the nitrogen {\it K} binding
energies for different geometries \cite{rempel09,esch99}. The structural sensitivity of such
reactions is fundamental to the understanding of heterogeneous catalysis.
Atomic {\it K}-edge spectroscopy may also prove critical to understanding a
non-molecular phase of solid nitrogen, identified at high pressures via optical
measurements, and likely related to an insulator-to-metal phase transition \cite{gonch00}.

These diverse applications typically rely on tabulations of photoionization cross
sections ({\it e.g.}, \cite{henke93}) determined from a combination of basic theory
and measurements of either molecular-phase or solid-phase N$_2$.
A lack of high-quality data for atomic nitrogen has clearly impeded
progress in some areas \cite{bmcl01}. Unfortunately no experiments have been reported
to date in the atomic nitrogen {\it K}-edge region.
The primary barrier to such experiments, of course, is the difficulty of producing a dense atomic-nitrogen
sample as a target. Experimental studies of atomic nitrogen in its ground state
$\rm 1s^22s^22p^3~^4S^o$ in the vicinity of the {\it K} edge may be further
hampered by both the presence of metastable states of the atomic species
as well as molecular resonances, such as $\rm 1s \rightarrow \pi^{\ast}$,
in the N$_2$ gas used as a precursor in the production of atomic nitrogen.

In this Letter, we report high-resolution {\it K}-edge spectra of gas-phase atomic nitrogen,
in combination with state-of-the art theoretical calculations, in order to provide benchmark
values for x-ray photoabsorption cross sections of this important species.
The measurements were performed in the 390-418 eV photon-energy range,
with a resolution of 60 meV, on the high-brightness undulator beamline 8.0 at the
Advanced Light Source (ALS) synchrotron radiation facility in Berkeley, California.
The atomic-nitrogen target was produced by partial dissociation of N$_2$ molecules in a microwave
discharge \cite{stolte97} and introduced to the interaction region with the x-ray beam.
Singly, doubly, and triply charged nitrogen cations were extracted from this region,
selected with a magnetic mass analyzer \cite{stolte08}, and their yields were
measured as a function of photon energy. Careful extraction of the ion yields
due to undissociated N$_2$ in the target permits determination of partial-ion
yields due solely to photoionization of atomic nitrogen. Fortunately, while
the intense N$_2$ $\rm 1s\rightarrow\pi^{\ast}$ molecular resonance is
present in this energy region, it does not overlap with any atomic-nitrogen
features, allowing easy discernment of the nitrogen
1s2s$^2$2p$^3$$n$p core-excited atomic resonances.

\begin{figure}[hbt]
\includegraphics[scale=1.0,width=6.25cm]{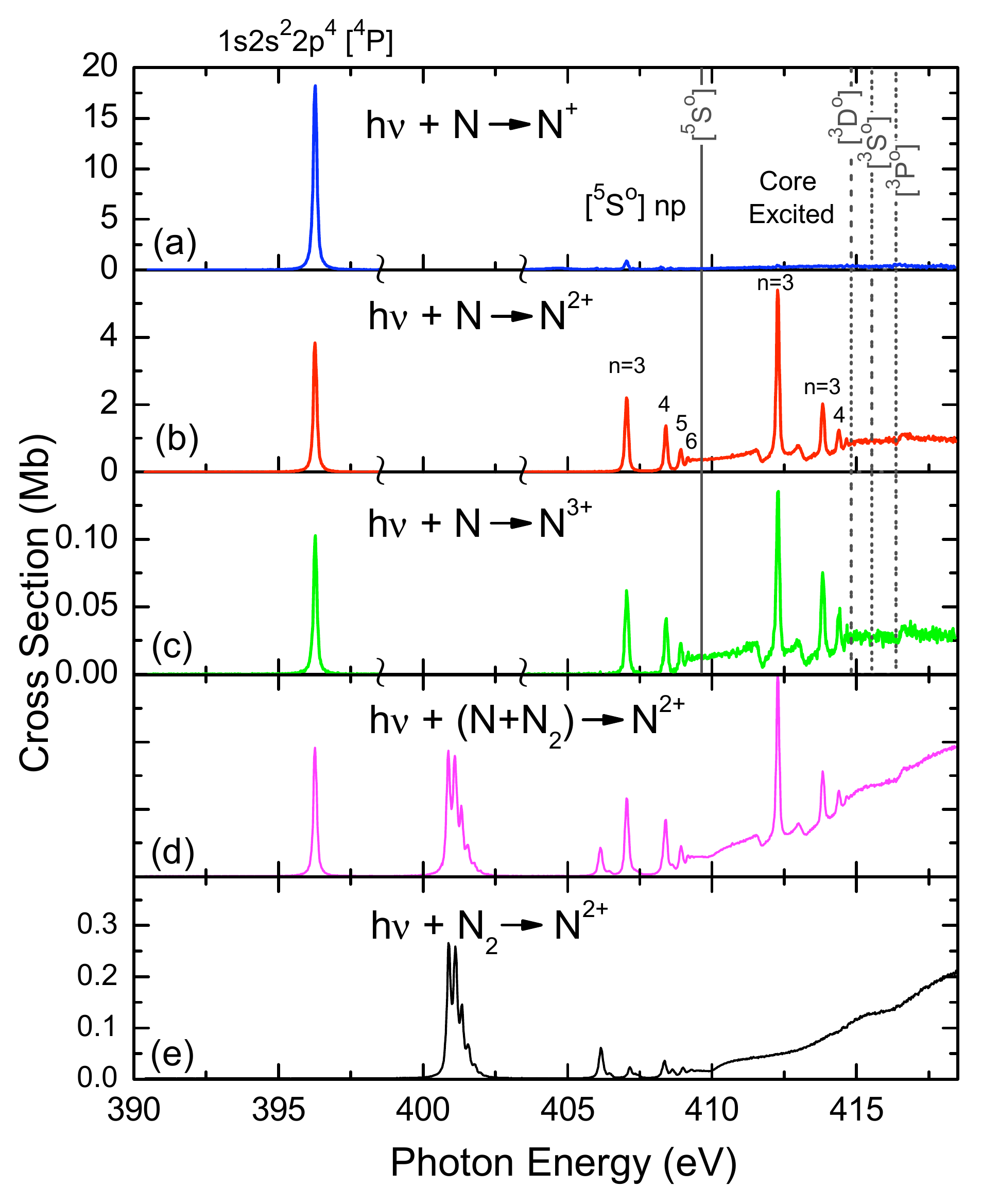}
\caption{\label{Fig1color} Partial cross sections for photoionization of atomic nitrogen 
					to (a) N$^+$, (b) N$^{2+}$, and (c) N$^{3+}$. 
					Figures (d) and (e)  are experimental results from a mixture of atomic/molecular 
					nitrogen and molecular nitrogen resulting in N$^{2+}$.}
\end{figure}

For the present measurements, gaseous N$_2$ was passed into a microwave cavity, and all products
of the nitrogen plasma then flowed into the interaction region after passing through an L-shaped Pyrex tube \cite{stolte97}.
It was found that collisions with the walls of this tube strongly quenched the nitrogen atoms created
in long-lived metastable states ($\rm 1s^22s^22p^3$[$\rm ^2D^o$,
$\rm ^2P^o$]) before they reached the interaction region, leaving only $\rm 1s^22s^22p^3$[$\rm ^4S^o$]
ground-state atoms and molecular N$_2$ to interact with the x-rays (examples are shown in Fig 1(d) and 1(e)).
To enhance the efficiency of the microwave discharge, a constant magnetic field, satisfying the electron-cyclotron resonance
condition, was superimposed on the microwave cavity perpendicular to the 2.45 GHz electric and magnetic fields.
The magnetic field aided in confinement and electron-cyclotron heating of the plasma, increasing the
dissociation fraction of N$_2$ \cite{geddes94} (which is intrinsically low due to the strength of the triple bond), to 4\% for the present experiments.
While dissociation fractions as high as 80\% for O$_2$ \cite{stolte97} and 20\% for C$\ell_2$ \cite{sam86}
have been achieved for other open-shell atoms, previous measurements of valence-shell photoionization
of atomic nitrogen achieved only about 1\% \cite{sam90}. Thermal dissociation of N$_2$, an
alternate technique, is even less efficient \cite{marcelo97}.

The partial-ion yields measured as described above include contributions from both ground-state
atomic nitrogen and molecular N$_2$, but which differ as a function of photon energy.
Since the spectroscopy of molecular N$_2$ is well known \cite{chen89,semen06},
its presence provides an internal energy calibration. The cross section, $\sigma^{q+}$ ${\rm (E)}$,
as a function of photon energy, E, for photoionization of atomic nitrogen to an ion of charge +$q$ can be obtained from \cite{sam85}:
\begin{equation}
\sigma^{q+} {\rm (E)} = {\rm C}_{q^+} ({\rm I}_{\rm on}^{q+} -  f  \times {\rm I}_{\rm off}^{q+}),
\label{eq:ion}
\end{equation}
where I$_{\rm on}^{q+}$, Fig 1(d), and I$_{\rm off}^{q+}$, Fig 1(e), are normalized ion yields measured as a function of
photon energy with the microwave discharge on or off, and C$_{q+}$ is a constant dependent on the
number density of nitrogen atoms and the ion-collection efficiency of the apparatus.
Absolute data for single and multiple photoionization of N$_2$ \cite{stolte98} were used to determine
values for the constants C$_{q+}$. The parameter $f$ = $(n {(\rm{N}_2^{\rm on})/} n (\rm{N}_2^{\rm off})$), with $n {(\rm{N}}_2^{\rm on})$
and $n (\rm{N}_2^{\rm off})$ being number densities of N$_2$ with the microwave discharge on or off, represents
the fraction of N$_2$ molecules that do not dissociate in the discharge. The value of $f$ is empirically chosen
to eliminate the molecular peaks from the measured ion yields via a weighted subtraction \cite{sam86,sam90}.
As noted above, the dissociation fraction, 1-$f$, was about 4\%. Finally, the collection efficiency for each
ion N$^{q+}$ produced by photoionization of atomic nitrogen was assumed equal to the collection
efficiency of the same ion generated in dissociative photoionization of N$_2$.

Figure 1 (a) -- (c) shows partial cross section spectra for single, double, and triple photoionization of atomic nitrogen
in the photon-energy range 390.0-418.5 eV. The energy region from 398.5-403.5 eV is intentionally
left blank due to the presence of spurious structure from the subtraction of two large numbers in the vicinity of the
$\rm 1s \rightarrow {\pi}^{\ast}$ molecular resonance. The strong peak in the cross section at 396.26 eV is due to 1s $\rightarrow$ 2p promotion from the
$\rm 1s^22s^22p^3[^{4}S^o]$ ground state to the lowest-lying core-excited $\rm 1s2s^22p^4[^4P]$ autoionizing
resonance state. At higher energies, the peaks correspond to transitions to $\rm 1s2s^22p^3[^{5}S^o]$n$p$ resonances
below the lowest {\it K}-edge threshold as well as
$\rm 1s2s^22p^3[^{3}D^o]$n$p$, $\rm 1s2s^22p^3[^{3}S^o]$n$p$, and $\rm 1s2s^22p^3[^{3}P^o]$n$p$
core-excited resonance states leading to higher-energy thresholds, as indicated in the figure and in Table I.
Similar to molecular photofragmentation, N$^+$ is the prevalent ion formed, and most pronounced on 
the first resonance, whereas the other charge states are stronger for the higher lying resonances. 
Although atomic and molecular nitrogen have different electronic structures, our observed 
natural line width of 111 meV for the $\rm 1s2s^22p^4[^4P]$ resonance is virtually identical 
to the 113 eV  previously observed for the N$_2$ (N 1s, $\nu$=0)$\rightarrow$($\pi_g$*,$\nu\,^{\prime}$) state \cite{kato}.

State-of-the-art {\it ab-initio} theoretical methods are routinely used now
\cite{bmcl01, fred04, scully05, garcia05, scully06, mueller09, witthoeft09, mueller10, fatih10} to study inner-shell atomic processes,
earlier studies were limited and primarily used the Hartree-Fock (HF) approximation \cite{rm79,yeh93,chant95}.
The present theoretical predictions of atomic-nitrogen photoabsorption cross sections
 used the R-matrix method \cite{rmat}: in this close-coupling approximation the core ion is
 represented by an N-electron system, and the total wavefunction expansion, $\Psi$(E), of the (N+1)-electron
 system for any symmetry $SL\pi$ is expressed as
\begin{equation}
\Psi({\rm E}) = A \sum_{i} \chi_{i}\theta_{i} + \sum_{j} c_{j} \Phi_{j},
\label{eq:rmat}
\end{equation}
\noindent
where $\chi_{i}$ is the core wavefunction in a specific state $S_iL_i\pi_i$ and $\theta_{i}$ is the wavefunction
 for the (N+1)th electron in a channel labeled as $S_iL_i\pi_ik_{i}^{2}\ell_i(SL\pi)$, with $k_{i}^{2}$ being its incident kinetic energy.
 The $\Phi_j$ are correlation wavefunctions of the (N+1)-electron system that account for short-range correlation and
 orthogonality between continuum and bound orbitals, with $c_j$ being the expansion coefficients.
  $\Psi({\rm E})$ is a bound (N+1)-electron wavefunction
 when the energy $\rm E < 0$, with all channels closed, and a continuum wavefunction when $\rm E > 0$, with some or all continuum channels open.

All of the (390) target levels in this work were represented by multi-configuration-interaction wavefunctions, 
obtained from $n=2$ physical and pseudo 3$\overline{\ell}$ orbitals of the residual N$^+$ ion core. 
The $LS$ coupling R-matrix cross-section calculations included both radiative and Auger damping.
The scattering wavefunctions were generated by allowing two-electron promotions out of the designated base configurations.
The resonance features were resolved in the scattering problem with a fine-energy mesh of 1.36 $\mu$eV (10$^{-7}$ Rydbergs).

Results for the autoionizing resonances are presented in Table I, indicating good agreement between experiment and theory.
Precise energy calibration was achieved by comparing positions of the atomic-nitrogen features
to resonances in the well-known molecular spectrum \cite{chen89}, specifically the narrow 3s$\sigma$ (406.150 eV) and
3p$\pi$ (407.115 eV) Rydberg transitions. The absolute energy scale was confirmed using the $v_0$ vibrational level of the
$\rm 1s \rightarrow {\pi}^{\ast}$ resonance at 400.868 eV \cite{chen89}. The atomic-nitrogen ion-yield spectra (Fig. 1)
were fitted to Voigt functions for the peaks and arctangent functions
 for the thresholds to obtain accurate peak positions.
 Experimentally, the lowest-energy 1s $\rightarrow$ $n$p Rydberg series, with a
 N$\rm {^+}[{^5S^o}]$ core, was found to converge to a binding energy of 409.64 $\pm$ 0.02 eV,
 based on the Rydberg formula. This threshold is 0.30 $\pm$ 0.02 eV lower than the accepted
 value for the atomic nitrogen 1s binding energy in N$_2$ \cite{chen89, Johan73}.
\begin{figure}[htb]
\includegraphics[scale=0.29,angle=0]{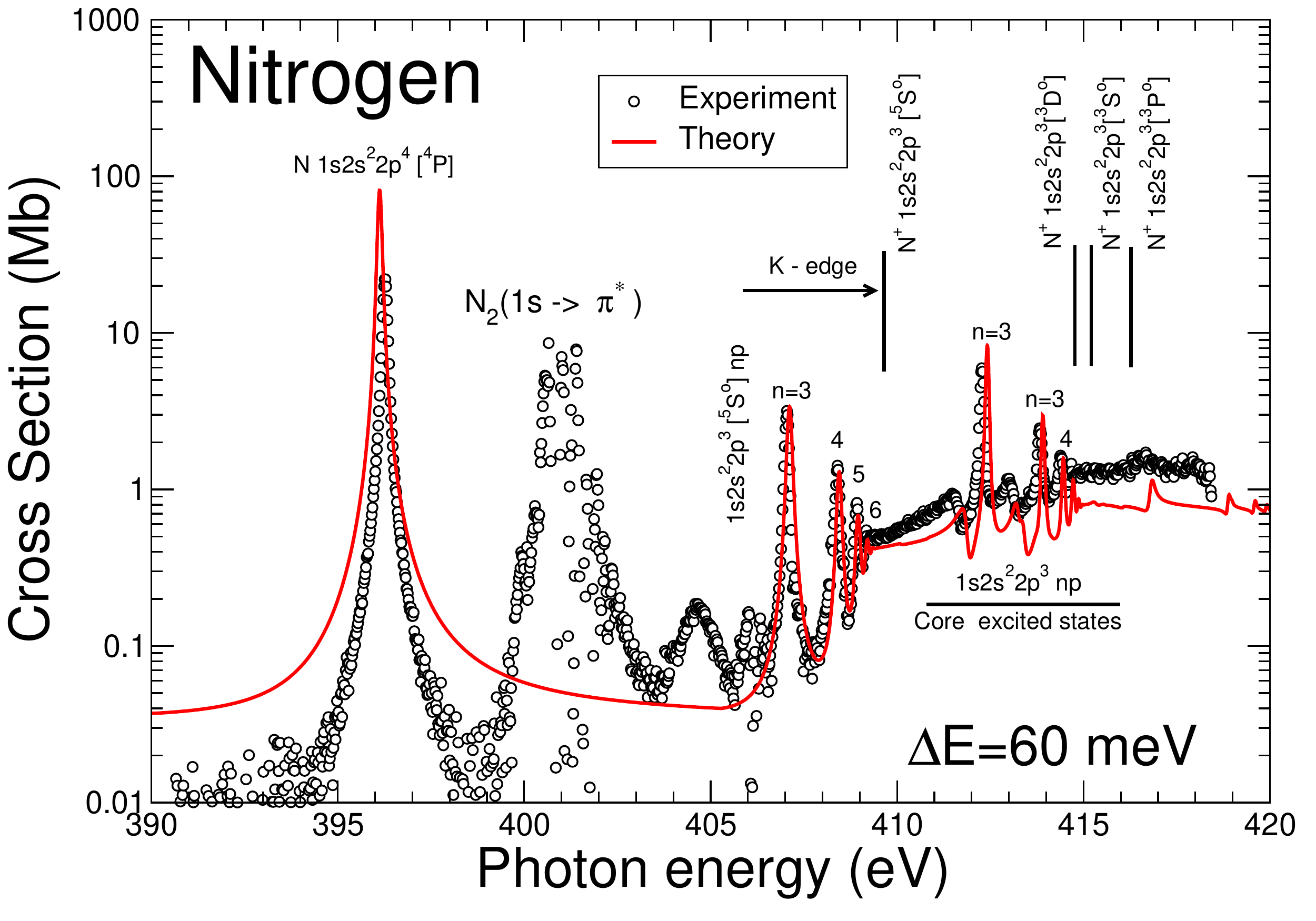}
\caption{\label{Fig2color}Atomic-nitrogen total photoionization cross section.
                                            Theoretical results were convoluted with a 60 meV FWHM Gaussian to simulate experiment.
				       Experimental results include molecular components between 400 and 406 eV.}
\end{figure}

The experimental cross sections for photoionization of atomic nitrogen to N$\rm ^{+}$, N$\rm ^{2+}$, and N$\rm ^{3+}$
ions in Fig. 1 were summed to obtain the total photoionization cross section, shown in Fig. 2 along with theoretical
predictions using the R-matrix method. Good agreement is found between theory and experiment for both
the resonance energies and absolute cross sections.
Observed differences in the background between experiment and theory are consistent with uncertainties resulting from the
experimental normalization procedure. Approximately 96\% of our partially dissociated target is composed of N$_2$ 
 and as we are subtracting in (1) two large numbers of comparable magnitude, the contribution from N$_2$ in I$_{\rm on}^{q+}$
is not completely equivalent to the room-temperature I$_{\rm off}^{q+}$,  the vibrational structure is different. 
This is clearly observed in Fig. 2 by the remaining $\pi$* resonance with the 
discrepancy manifesting itself more with increasing photon energy.
   Our results demonstrate strong resonant features dominate
the photoionization cross section of atomic nitrogen in the vicinity of the {\it K} edge, the atomic resonances differing
significantly from those in molecular nitrogen which are absent in existing tabulations.  Rydberg states
above the first ionization threshold at 409.64 eV show clear Fano-Beutler profiles, indicating strong interferences
between the core-excited intermediate resonance states and the open continua.

\begin{table*}[thb]
\caption{\label{Auto}Atomic nitrogen $\rm 1s2s^22p^3~np~^4P$ core-excited autoionizing resonances
               converging to the $\rm ^5S^o$, $\rm ^3D^o$, $\rm ^3S^o$ and $\rm ^3P^o$ series limits.
               Results are given for the resonance positions (eV), natural linewidths $\Gamma$ at FWHM (meV) and
               quantum defects $\mu$. Experimental linewidths were extracted from the ion-yield measurements in Fig. 1
               by fitting Voigt profiles to the peaks assuming a Gaussian instrumental contribution of 60 meV for each peak.}
\begin{ruledtabular}
\begin{tabular}{ccccccc}
%Resonances              & Energy (eV)   	& Energy (eV) 	&$\Gamma$(meV)	&$\Gamma$(meV)	& $\mu$  & $\mu$   \\
Resonances                & Energy (Expt)   & Energy (Theory) &$\Gamma$ (Expt)	&$\Gamma$ (Theory)& $\mu$ (Expt)  & $\mu$ (Theory)   \\
% (label)                              & (Expt)       		& (Theory)    	&   (Expt)    		& (Theory)    		& (Expt)	  	&(Theory) \\
\hline
\\
$\rm 1s2s^22p^4$~~~~~~     &396.27 $\pm$ 0.02    &396.18             	&111 $\pm$ 10     	&109        		 &0.99 $\pm$ 0.02     &0.99\\
$\rm 1s2s^22p^3[^5S^o]3p$ &407.05 $\pm$ 0.02    &407.08             	&~~99 $\pm$ 20     	&107                   &0.71 $\pm$ 0.02	&0.70\\
$\rm 1s2s^22p^3[^5S^o]4p$ &408.40 $\pm$ 0.02    &408.46             &~~89 $\pm$ 20     	&106                   &0.69 $\pm$ 0.02	&0.61\\
$\rm 1s2s^22p^3[^5S^o]5p$ &408.92 $\pm$ 0.03    &408.94             &~~95 $\pm$ 10       &~98               	 &0.65 $\pm$ 0.03	&0.59\\
$\rm 1s2s^22p^3[^5S^o]6p$ &409.16 $\pm$ 0.04    &409.18             	&~~80 $\pm$ 10       &~80                   &0.68 $\pm$ 0.04	&0.56\\
$\rm ^5S^o$ series limit         &409.64 $\pm$ 0.02    & ~	            	&~                        	&~                   	 & ~     			&  \\
--                        ~~~ ~~~          & --                                  & --            	&~ --                       	&~--                      & ~ --      			&~ --  \\
$\rm 1s2s^22p^3[^3D^o]3p$ &412.28 $\pm$ 0.02    &412.42            	&~~83 $\pm$ 10     	&~80                    &0.69 $\pm$ 0.02	& 0.62\\
$\rm ^3D^o$ series limit         &414.82 $\pm$ 0.05    & ~	                   &~                        	&~                   	 & ~     			&       \\
--                        ~~~ ~~~          & --                                  & --            	&~ --                       	&~--                      & ~ --      			&~ --  \\
$\rm 1s2s^22p^3[^3S^o]3p$ &412.96 $\pm$ 0.06  	&413.14             &~237 $\pm$ 20	&235              	  &0.70 $\pm$ 0.06	& 0.62\\
$\rm 1s2s^22p^3[^3S^o]4p$ &414.38 $\pm$ 0.03    &414.45             &~116 $\pm$ 10       &107               	  &0.56 $\pm$ 0.03	& 0.45\\
$\rm ^3S^o$ series limit         &415.53 $\pm$ 0.05  	& ~	   		&~   				&~  			  & ~     			&  \\
--                        ~~~ ~~~          & --                                  & --            		&~ --                       	&~--                      & ~--      			&  \\
$\rm 1s2s^22p^3[^3P^o]3p$ &413.83 $\pm$ 0.02    &413.91             &~100 $\pm$ 10     	&108                     &0.68 $\pm$ 0.02	& 0.64\\
$\rm ^3P^o$ series limit         &416.36 $\pm$ 0.05	& ~	                   &~              		&~                   	  & ~     			&  \\
\end{tabular}
\end{ruledtabular}
\end{table*}
%+++++++++++++++++++++++++++++++++++++++++++++++++++++++++++++++++++++++++++++
%
The experiments were supported by NSF, DOE, the DOE Facilities Initiative, Nevada DOE EPSCoR, and CNPq (Brazil).
The theoretical work was supported by NSF and NASA grants to ITAMP at the Harvard-Smithsonian Center for Astrophysics
and the Smithsonian Astrophysical Observatory. The computational work was done at the National Energy Research
Scientific Computing Center in Oakland, CA and on the Tera-grid at the National Institute for Computational
Sciences (NICS), in Knoxville, TN, which is supported in part by NSF.
%
%+++++++++++++++++++++++++++++++++++++++++++++++++++++++++++++++++++++++++++++
%
%   Reference section now follows
%
%   Delete or change fake bibitem. delete next three
%   lines and directly read in your .bbl file if you use bibtex.
%
%+++++++++++++++++++++++++++++++++++++++++++++++++++++++++++++++++++++++++++++
%

\end{document}